\documentclass[preprint,superscriptaddress,nofootinbib]{revtex4}
\usepackage[dvips]{graphicx}
\usepackage[usenames]{color}
\usepackage{setspace}
\usepackage{amsmath}
\usepackage{amssymb}
\usepackage{amsthm}

\begin{document}
%%%%%%%%%%%%%%%%%%%%%%%%%%%%%%%%%%
%%%%%%%%%%% Title page %%%%%%%%%%%
%%%%%%%%%%%%%%%%%%%%%%%%%%%%%%%%%%

  \preprint{UT-HET 101}

  \date{\today}

%%%%%%%%%%%%%%%%%%%%%%%%%
  \title{Indirect reach of heavy MSSM Higgs
bosons by precision
measurements at future lepton colliders}
%%%%%%%%%%%%%%%%%%%%%%%%%
  \author{Mitsuru~Kakizaki}
  \email{kakizaki@sci.u-toyama.ac.jp}
  \affiliation{
  Department of Physics,
  University of Toyama, Toyama 930-8555, Japan
  }
%%%%%%%%%%%%%%%%%%%%%%%%%
  \author{Shinya~Kanemura}
  \email{kanemu@sci.u-toyama.ac.jp}
  \affiliation{
  Department of Physics,
  University of Toyama, Toyama 930-8555, Japan
  }
%%%%%%%%%%%%%%%%%%%%%%%%
  \author{Mariko~Kikuchi} 
  \email{kikuchi@jodo.sci.u-toyama.ac.jp}
  \affiliation{
  Department of Physics,
  University of Toyama, Toyama 930-8555, Japan
  }
%%%%%%%%%%%%%%%%%%%%%%%%
  \author{Toshinori~Matsui} 
  \email{matsui@jodo.sci.u-toyama.ac.jp}
  \affiliation{
  Department of Physics,
  University of Toyama, Toyama 930-8555, Japan
  }
%%%%%%%%%%%%%%%%%%%%%%%%
  \author{Hiroshi~Yokoya} 
  \email{hyokoya@sci.u-toyama.ac.jp}
  \altaffiliation[Present address: ]{Theory Center, KEK}
  \affiliation{
  Department of Physics,
  University of Toyama, Toyama 930-8555, Japan
  }
%%%%%%%%%%%%%%%%%%%%%%%%
  \begin{abstract}

In the Minimal Supersymmetric Standard Model (MSSM), the bottom Yukawa
coupling of the Higgs boson can considerably deviate from its Standard
Model prediction due to non-decoupling effects.  
We point out that
the ratio of the Higgs boson decay branching fraction to a bottom quark pair
and that to a $W$-boson pair from the
same production channel is particularly sensitive to large additional
MSSM Higgs boson mass regions at future electron-positron colliders.
Based on this precision measurement, we explicitly show the
indirect discovery reach of the additional Higgs bosons according to
planned programs of the International Linear Collider.

  \end{abstract}
%%%%%%%%%%%%%%%%%%%%%%%%

\maketitle

%\cleardoublepage
%\tableofcontents

\cleardoublepage
%\pagenumbering{arabic}

%%%%%%%%%%%%%%%%%%%%%%%%%%%%%
%\section{Introduction}
%%%%%%%%%%%%%%%%%%%%%%%%%%%%%

In July 2012, discovery of a new particle with a mass of 125 GeV was
announced by the ATLAS and CMS collaborations at the CERN Large Hadron
Collider (LHC) \cite{LHC}.  Properties of the discovered particle have
turned out to be consistent with those of the Higgs boson of the
Standard Model (SM) within the error.  The SM has been experimentally
confirmed as a low-energy effective theory that consistently describes
every phenomenon at energy scales below ${\cal O}(100)$ GeV.
Although the SM is very successful, several phenomena which
cannot be explained in the framework of the SM have been known, such as
neutrino oscillations, the abundance of dark
matter and baryon asymmetry of the Universe.  
In order to account for such strong evidence we need to go beyond
the SM.

It should be emphasized that many new physics models which can explain the
new phenomena mentioned above demand extension of the Higgs sector.  Given
such a non-minimal Higgs sector, a SM-like Higgs boson with
a mass of 125 GeV is not exactly
the SM one, and has different characteristic properties from their SM
predictions.  Therefore, it is plausible that hints of new physics are
obtained by investigating properties of the discovered Higgs boson.
In this viewpoint, the discovered Higgs boson is a window to new
physics.

Indeed, one of the best strategies for exploring new physics is to
investigate its effects that appear in the couplings of the discovered
Higgs boson to SM particles.  Since the magnitudes of these Higgs
boson couplings depend strongly on masses and couplings of particles
yet to be discovered, fingerprinting of the Higgs boson couplings is
useful in determining the energy scale of new physics and further in
distinguishing new physics models behind.  At the LHC, deviations of
the Higgs boson couplings from their SM predictions are currently
constrained typically with 10\% \cite{LHC-kappa}.  At future
electron-positron colliders such as the International Linear Collider
(ILC) \cite{ILC}, the Compact LInear Collider (CLIC) \cite{CLIC} and
the Future Circular Collier of electrons and positrons (FCC-ee)
\cite{FCC-ee}, accuracies for measurements of various Higgs boson
couplings will be significantly improved to a percent level or better
\cite{Peskin:2012we,ILC-TDR2,ILCHiggsWhitePaper}.  Loop corrections
from new particles also give characteristic deviation patterns in
these couplings.  Precision measurements at future collider
experiments can differentiate such small deviations.  Therefore, it is
crucial to perform fingerprinting of the deviation pattern of the
Higgs boson couplings based on precise computations including
radiative corrections in beyond-the-SM (BSM) models for estimating the
scale of new physics indirectly.

Among ever proposed BSM models, supersymmetric (SUSY) extension of the
SM is one of the excellent paradigms \cite{SUSY}.  Introducing
superparticles whose spins are different from the SM counterparts by
one half, an unnatural cancellation between the bare mass squared of
the Higgs boson and its quadratic ultraviolet divergences is avoided.
By imposing the $R$-parity, the lightest SUSY particle (LSP) is
stabilized and thus can be an excellent candidate for dark matter.
The $SU(3)_C$, $SU(2)_L$ and $U(1)_Y$ gauge coupling constants can be
naturally unified at a high energy scale, inspiring many ideas about
supersymmetric grand unified theories.  One of the intriguing features
is that the Higgs sector of SUSY models must be extended to cancel the
gauge anomaly.  Even in the Minimal Supersymmetric Standard Model
(MSSM), two Higgs doublets $H_1^{}$, $H_2^{}$ with opposite
hypercharges are required.  As a consequence, the coupling strengths
of the discovered Higgs boson with a mass of 125 GeV to SM particles
deviate from their SM predictions, and have the same structure as in the
type-II two Higgs doublet model at the tree level.  At the loop level,
however, superparticles radiatively affect the Higgs boson coupling
strengths, giving rise to characteristic deviations patterns different
from other models.  In particular, it has been known that significant
deviations of the bottom and tau Yukawa couplings induced by
non-holomorphic radiative corrections do not decouple even in the
large superparticle mass limit
\cite{Hall:1993gn,Hempfling:1993kv,Carena:1994bv,Carena:1998gk,Babu:1998er,Eberl:1999he,Carena:1999py,Haber:2007dj,Cahill-Rowley:2014wba,Endo:2015oia}.
Due to the non-decoupling effects, the indirect reach of the
additional Higgs boson mass scale through precision measurements of
the bottom/tau Yukawa coupling can be much higher than the direct
reach at the LHC and its luminosity-upgraded version.

In this letter, we reanalyze the deviation of the Higgs boson
couplings enhanced by the loop-induced non-decoupling effects from a
difference perspective.  In sharp contrast to recent relevant works
\cite{Cahill-Rowley:2014wba,Endo:2015oia}, we utilize the ratio of
${\rm Br}(h\to b\bar{b})$ to ${\rm Br}(h\to WW)$ from the same
production channel $e^+ e^- \to \nu \bar{\nu} h$ as a key observable.
Since this combination cancels out the uncertainties related to the
Higgs boson production cross section and total decay rate, the
indirect additional Higgs boson reach obtained through this method is
more expanded than using single coupling accuracies.  Since the cross
section times branching ratios of these decay modes are particularly
accurately measurable, our quick method is not inferior to global
analysis.  As an example, we assume expected accuracies at the ILC
\cite{ILCHiggsWhitePaper}, whose running programs are under
discussion.  Our goal is to explicitly show the indirect discovery
reach of the additional MSSM Higgs bosons through the measurement of
the ratio ${\rm Br}(h\to b\bar{b})/{\rm Br}(h\to WW)$ at several
planned ILC programs.  Such an analysis plays an important role in
properly assessing the capability of future electron-positron
colliders for exploring new physics effects.

%%%%%%%%%%%%%%%%%%%%%%%%%%%%%
%\section{Higgs sector}
%%%%%%%%%%%%%%%%%%%%%%%%%%%%%

We start briefly reviewing the MSSM Higgs sector, which consists of
two Higgs doublet scalars.  The electroweak symmetry is broken when
the neutral components of $H_1^{}$ and $H_2^{}$ develops vacuum
expectation values (VEVs). Down-type quarks and charged leptons
(up-type quarks) acquire their masses from the VEV $\langle H_1^0
\rangle = v_1/\sqrt{2}$ ($\langle H_2^0 \rangle= v_2/\sqrt{2}$).  The
ratio $\tan \beta = v_2/v_1$ is an important quantity on which many
observables depend, and the sum of the squares is given by
$v^2=v_1^2+v_2^2=(246~{\rm GeV})^2$.  After the electroweak symmetry
breaking, there are two $CP$-even Higgs bosons $h$ and $H$, a $CP$-odd
Higgs boson $A$ and charged Higgs bosons $H^{\pm}$ as physical
particles.  The lighter $CP$-even Higgs boson $h$ is identified as the
discovered SM-like Higgs boson at the LHC.  At the tree level, its
predicted mass is smaller than the mass of the $Z$-boson as the Higgs
self-interactions are given by $SU(2)_L$ and $U(1)_Y$ gauge couplings.
However, the radiative correction significantly lifts the mass of the
Higgs boson and can account for its observed mass of 125 GeV although
superparticles are required to have a few TeV masses
\cite{MSSMHiggsMass}.  The mixing angle $\alpha$ between the two
$CP$-even Higgs bosons is also affected by the radiative correction.
For detailed discussion about the radiative corrections to the masses
of the $CP$-even Higgs bosons and its mixing angle, see, for example,
Ref.~\cite{Djouadi:2005gj}.  By virtue of supersymmetry, $H$, $A$ and
$H^{\pm}$ are degenerate in mass in the large mass limit.  We consider
the $CP$-odd Higgs boson mass $m_A$ as a measure of the mass scale of
these heavy MSSM Higgs bosons.

%%%%%%%%%%%%%%%%%%%%%%%%%%%%%%%%%
%\section{Bottom Yukawa Coupling}
%%%%%%%%%%%%%%%%%%%%%%%%%%%%%%%%%

Let us discuss non-decoupling effects of radiative corrections to Yukawa
couplings in the MSSM.  
In this letter we focus on the bottom Yukawa
coupling because it receives large radiative corrections from the
strong coupling and the top Yukawa coupling as well as future linear
colliders are capable of measuring the $h\to b \bar{b}$ decay rate with
considerable accuracy.
In SUSY models, the holomorphy restricts possible form of the
superpotential.  
As a consequence, the bottom quark does not
couple to $H_2$ at the tree level.  However, effects of SUSY breaking
induce a non-holomorphic bottom Yukawa coupling $\Delta \lambda_b$ radiatively
in addition to holomorphic radiative corrections 
$\delta \lambda_b$ \cite{Babu:1998er}:
\begin{eqnarray}
  - {\cal L}_b = (\lambda_b + \delta \lambda_b ) \bar{b}_R H_1 Q_L
  + \Delta \lambda_b \bar{b}_R Q_L H_2^*\, ,
\end{eqnarray}
where $b_R$ and $Q_L$ denote the right-handed bottom quark and
the third generation quark doublet, respectively.
Notice that such non-holomorphic couplings are absent in the type-II
two-Higgs-doublet model due to the hypothetical $Z_2$ symmetry.
The mass of the bottom quark is also shifted
by the radiative corrections as
\begin{eqnarray}
  m_b = \frac{\lambda_b v}{\sqrt{2}} \cos \beta (1 + \Delta_b)\, ,
\end{eqnarray}
where
\begin{eqnarray}
  \Delta_b \equiv \frac{\delta \lambda_b}{\lambda_b} + \frac{\Delta \lambda_b}{\lambda_b} \tan \beta\, .
\end{eqnarray}
The coupling of the SM-like Higgs boson to the bottom quark is obtained as
\begin{eqnarray}
  g_{hb\bar{b}}^{} = \frac{g m_b}{2 m_W}
  \frac{\sin \alpha}{\cos \beta}
  \left[ 1 + \frac{1}{1+ \Delta_b} \left( \frac{\delta \lambda_b}{\lambda_b}
      - \Delta_b \right) (1 + \cot \alpha \cot \beta) \right]\, .
\end{eqnarray}

Dominant contributions to the non-holomorphic coupling stem
from sbottom-gluino and stop-chargino loop diagrams and are given by
\begin{eqnarray} \label{eq:Deltab}
  \Delta_b \simeq \left( \frac{2 \alpha_s}{3 \pi}
\frac{\mu M_3}{m_{\rm SUSY}^2} + \frac{\lambda_t^2}{16 \pi^2}
\frac{\mu A_t}{m_{\rm SUSY}^2} \right) \tan \beta\, ,
\end{eqnarray}
where $\mu$, $M_3$ and $A_t$ are the higgsino mass, the
gluino mass and the trilinear stop $A$-parameter, respectively.
The typical mass
scale of the superparticles in the loop diagrams is collectively
denoted by $m_{\rm SUSY}^{}$.  Notice that there are two mass scales
characterizing effects of new physics in the MSSM.  One is the mass
scale of superparticles $m_{\rm SUSY}$, and the other is the mass of
the $CP$-odd Higgs boson $m_A$.  In the limit where superparticles are
heavy $\mu \sim M_3 \sim A_t \sim m_{\rm SUSY} \gg m_Z$ with fixed
$m_A$, the bottom Yukawa coupling is approximately given by
\begin{eqnarray} \label{eq:ghbbapp}
  g_{hb\bar{b}}^{} \simeq \frac{g m_b}{2 m_W}
  \frac{\sin \alpha}{\cos \beta}
  \left[ 1 - \Delta_b (1 + \cot \alpha \cot \beta) \right]\, .
\end{eqnarray}
Therefore, SUSY loop corrections do not decouple for small $m_A$, and
are enhanced for large $\tan \beta$.  On the other hand, in the limit
where the mass of the $CP$-odd Higgs boson is large, we obtain
\begin{eqnarray}
  1 + \cot \alpha \cot \beta = - \frac{2m_Z^2}{m_A^2 }\cos 2\beta 
  + {\cal O}\left(\frac{m_Z^4}{m_A^4} \right)\, ,
\end{eqnarray}
and thus SUSY radiative corrections decouple.  It should be noticed
that two-Higgs-doublet models can also produce non-decoupling effect
on the Higgs boson couplings, which originates from extra Higgs boson
loop diagrams
\cite{hhh_2HDM,Arhrib:2003ph,h_2HDM,Osland:2008aw,Wu:2015nba}.

%%%%%%%%%%%%%%%%%%%%%%%%%%%%%
%\section{Analysis}
%%%%%%%%%%%%%%%%%%%%%%%%%%%%%

We here describe the method for our numerical analysis.  We utilize
the public code FeynHiggs2.10.3 to compute mass and mixing parameters
as well as decay branching fractions in the MSSM Higgs sector
\cite{FeynHiggs} \footnote{Changes in the recently updated version
  FeynHiggs2.11.0 are not relevant to our analysis.}.  The masses of
the $CP$-even Higgs bosons and the mixing angle are computed at the
two-loop level in the on-shell scheme.  Branching ratios of the
SM-like Higgs boson are computed at the one-loop level including QED
and QCD corrections.  We perform a random scan of MSSM parameters
avoiding tachyonic scalar masses.  To circumvent large $CP$-violating
processes, we constrain our analysis to real MSSM parameters.  For
simplicity, we assume that soft SUSY breaking $A$-parameters are given
by the product of the corresponding Yukawa coupling constant
$\lambda_f$ and a common $A_0$ value as $A_f = \lambda_f A_0$.  We
consider the mass range of electroweak (colored) superparticles up to
around 1000 GeV (4000 GeV) to cover less tuned parameter sets which
evade superparticle searches at the future LHC experiment.  Since the
non-decoupling effects depend on ratios of superparticle masses, our
results about the Higgs boson couplings are not substantially changed
even for heavier superparticles.  If $|A_t|$ is significantly larger
than the stop masses, there appear a deeper charge/color breaking
vacuum than our electroweak vacuum.  To avoid complexity of the
discussion of the meta-stability bound, we constrain $|A_0|$ to be
smaller than three times the geometric average of the soft stop
masses.  The scan bounds of the MSSM parameters are listed in
Tab. \ref{tab:parameters}.  Since radiative corrections from the first
two generation sleptons and squarks to the Higgs sector are small, we
set the soft masses of these sleptons and squarks to $1000$ GeV and
$4000$ GeV, respectively, instead of performing a random scan.  This
degeneracy also suppresses unwanted flavor chaining neutral currents
and lepton flavor violating processes adequately.  Although the
measured mass of the Higgs boson is now fixed to $m_h= 125~{\rm GeV}$
with the error less than 1 GeV \cite{Aad:2015zhl}, we consider
$122~{\rm GeV} <m_h^{} < 128~{\rm GeV}$ as an allowed region for its
mass.  This is a judicious choice to take uncertainties from
renormalization scheme dependence \cite{MSSMHiggsMassRen} and higher
order contributions to the mass of the Higgs boson into account
\cite{Higgs3}.  From the cosmological viewpoint, it is assumed that
the lightest neutralino is the LSP and a dark matter candidate.
However, the relic density constraint is relaxed in order to allow for
any non-standard cosmological scenario.  As for the other experimental
constraints, we impose superparticle mass bounds obtained at the LEP
\cite{LEP} and the LHC \cite{LHC-BSM}.  The mass bounds we employ are
summarized in Tab. \ref{tab:constraints}.  The flavor violating decays
$B_s \to \mu^+ \mu^-$ and $b\to s \gamma$ mediated by the additional
Higgs bosons could also give constraints on the SUSY parameter space,
as studied in \cite{Endo:2015oia}.  However, such processes can be
canceled by contributions from squark sector parameters without
changing our results.  Therefore, we neglect these flavor constrains
in our analysis.  The deviation of the anomalous magnetic moment of
the muon from its SM prediction can be explained by adjusting the
masses of the second generation sleptons, which is beyond the scope of
this letter.  Since our analysis is confined to low-energy SUSY
scenarios, we require $\lambda_b$ to be perturbative at the
electroweak scale.

% Scanned parameter
\begin{table}
  \begin{center}
    \caption{Scan bounds on MSSM parameters.}
    \label{tab:parameters}
    \begin{tabular}{|c|c|}
    \hline
    Parameter & Scan bounds \\ \hline \hline
    $m_A^{}$ & [200 GeV, 3000 GeV] \\ \hline
    $\tan \beta$ & [1, 60] \\ \hline
    $\widetilde{m}_{L_3,E_3}^{}$ &  [100 GeV, 1000 GeV] \\ \hline
    $\widetilde{m}_{Q_3,U_3,D_3}^{}$ & [500 GeV, 4000 GeV] \\ \hline
    $|A_0^{}|$ & [0 GeV, $3(\widetilde{m}_{Q_3}\widetilde{m}_{U_3})^{1/2}$] \\ \hline
    $|\mu|$ & [100 GeV, 1000 GeV] \\ \hline
    $|M_1^{}|$ & [100 GeV, 1000 GeV] \\ \hline
    $|M_2^{}|$ & [100 GeV, 1000 GeV] \\ \hline
    $M_3^{}$ & [1400 GeV, 4000 GeV] \\ \hline
  \end{tabular}
  \end{center}
\end{table}

% Constraints
\begin{table}
  \begin{center}
    \caption{Constraints on superparticle masses.}
    \label{tab:constraints}
  \begin{tabular}{|c|c|}
    \hline
    Observables & Constraints \\ \hline \hline
    $m_h^{}$ & [122 GeV, 128 GeV]  \\ \hline
    $m_{\widetilde{g}}^{}$ & $>$ 1400 GeV \\ \hline
    $m_{\widetilde{\chi}_1^0}^{}$ & $>$ 90 GeV \\ \hline
    $m_{\widetilde{\chi}_1^\pm}^{}$ & $>$ 400 GeV \\ \hline
    $m_{\widetilde{t}_1}^{}$ & $>$ 640 GeV \\ \hline
    $m_{\widetilde{b}_1}^{}$ & $>$ 620 GeV \\ \hline
    $m_{\widetilde{\tau}_1}^{}$ & $>$ 90 GeV \\ \hline
  \end{tabular}
  \end{center}
\end{table}

%%%%%%%%%%%%%%%%%%%%%%%%%%%%%
%\section{Results}
%%%%%%%%%%%%%%%%%%%%%%%%%%%%%

Our goal is to show to what extent large $m_A$ scenarios with moderate
SUSY parameter choices can be surveyed through the precision
measurements of the cross section times branching ratio measurements
of the $h\to b \bar{b}$ and $h \to W W$ modes which follow the
production $e^+ e^- \to \nu \bar{\nu}h$.  To this end, we introduce
the following double ratio of the branching fractions of $h \to b
\bar{b}$ and that of $h \to WW$ modes:
\begin{eqnarray}
  R = \left. \frac{{\rm Br}(h \to WW)}{{\rm Br}(h \to b\bar{b})} \right|_{\rm MSSM}
      \bigg/ \left. \frac{{\rm Br}(h \to WW)}{{\rm Br}(h \to b\bar{b})} \right|_{\rm SM}\, .
\end{eqnarray}
Its tree level expression is given by
\begin{eqnarray} \label{eq:Rtree}
R_{\rm tree}^{} = \frac{1}{[1 - \tan \beta \cot (\beta - \alpha)]^2}\, ,
\end{eqnarray}
where $\cot(\beta-\alpha)$ is given as a function of $m_A$ and $m_Z$.
As is evident, any deviation from the SM prediction $R = 1$ exhibits
the existence of some new physics beyond the SM.  By taking the ratio
of the cross section times branching ratios for $b\bar{b}$ and $WW$
modes in the same production mechanism, uncertainties that originate
from the production cross section and the total decay rate of the
Higgs boson are canceled out.  Consequently, constraining the model
parameter space using the accuracy of the ratio of the decay branching
fraction $R$ is more powerful than using the converted accuracy of the
$hb\bar b$ coupling.  The expected accuracies for cross section times
branching ratio measurements of the SM-like Higgs boson through the
$\nu\bar{\nu}h$ mode at several ILC stages with the polarization
options $(P_{e^-}, P_{e^+})=(-0.8,+0.3)$ and $(P_{e^-},
P_{e^+})=(-0.8,+0.2)$ are listed in Tab. 5.4 and 5.5 of
Ref. \cite{ILCHiggsWhitePaper}.  Sensitivities for $R$ are computed
based on these tables and summarized in Tab. \ref{tab:R}.
For example, for $\sqrt{s}=500$~GeV and $L=500$~fb$^{-1}$,
the accuracy of the ratio at 95\% CL is estimated as
$\sqrt{(0.7\%)^2+(2.4\%)^2}\times2 = 5\%$,
while the accuracy of ${\rm Br}(h\to b\bar b)$ from the coupling measurement
itself is
at best $\simeq 6.4\%$ ignoring systematic uncertainties (see Table 6.1 in
Ref.[8]).

\begin{table}
  \begin{center}
    \caption{Expected accuracy for $R$ at planned stages of the ILC
      \cite{ILCHiggsWhitePaper}.}
  \label{tab:R}
  \begin{tabular}{|l|c|c|}
    \hline
    CM energy and luminosity & $95\%{\rm CL}$ & $5 \sigma$ \\ \hline \hline
    $\sqrt{s}=500~{\rm GeV}$, $L=500 ~{\rm fb}^{-1}$ & 0.050 & 0.125 \\ \hline
    $\sqrt{s}=500~{\rm GeV}$, $L=1600 ~{\rm fb}^{-1}$ & 0.027 & 0.068 \\ \hline
    $\sqrt{s}=1~{\rm TeV}$, $L=2500 ~{\rm fb}^{-1}$  & 0.021 & 0.052 \\ \hline
  \end{tabular}
  \end{center}
\end{table}

\begin{figure}[t]
\centering
\includegraphics{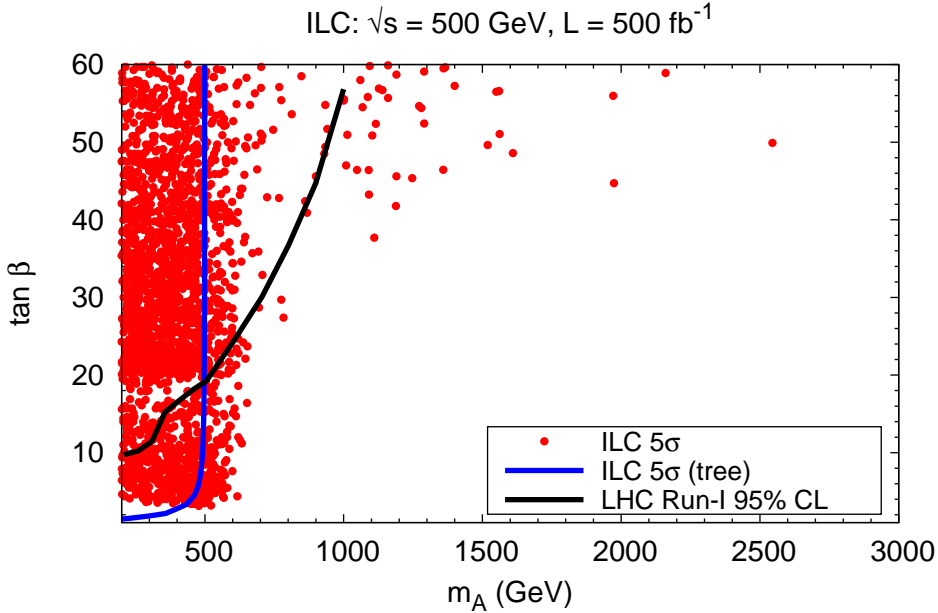}
\caption{\footnotesize Indirect $5 \sigma$ discovery reach of the
  heavy Higgs bosons through the measurement of $R$ at the ILC stage
  with $\sqrt{s}=500~{\rm GeV}$ and $L=500~{\rm fb}^{-1}$ in the
  ($m_A, \tan \beta$) plane (red points).  For comparison, the
  indirect $5 \sigma$ reach based on the tree-level results is
  delineated with blue line.  The 95\% exclusion limit obtained at the
  LHC Run-I through $H/A\to \tau^+ \tau^-$ searches is delineated with
  black line.}
\label{fig:ILC500}
\end{figure}

In Fig.~\ref{fig:ILC500}, the indirect $5\sigma$ discovery reach of
the MSSM heavy Higgs bosons through the measurement of $R$ at the ILC
with the center-of-mass (CM) energy $\sqrt{s}=500$ GeV and the
luminosity $L=500 ~{\rm fb}^{-1}$ is shown in the ($m_A, \tan \beta$)
plane.  MSSM scenarios which predict more than $5 \sigma$ deviation of
$R$ are plotted with the red points.  The blue line is the
corresponding tree level prediction based on $R_{\rm tree}$.  The
indirect reach of the MSSM heavy Higgs boson mass can be significantly
extended due to the non-decoupling property of the non-holomorphic
radiative corrections, compared to the tree level results.  When the
relative sign of $\mu$ and $M_3$ ($A_t$) is minus, the deviation of
$R$ from unity is strongly enhanced and the correction is proportional
to $\tan \beta$ (see Eqs.(\ref{eq:Deltab}) and (\ref{eq:ghbbapp})).
The solid (black) line shows the 95\% exclusion limit obtained at the
LHC Run-I through direct $H/A \to \tau^+ \tau^-$ searches
\cite{LHC-Htau}.  At the LHC Run-II with the luminosity of $L=10~{\rm
  fb}^{-1}$ accumulated, it is expected to exclude up to
$m_A=1000~{\rm GeV}$ for $\tan \beta=10$ in view of the increase of
the production cross section \cite{Morinaga}.  The existence of the
heavy Higgs boson can be searched also through the $H/A \to \mu^+
\mu^-$ decay modes at the LHC \cite{LHC-Hmu}.  Its expected $5\sigma$
discovery reach with $L=3000~{\rm fb}^{-1}$ almost overlaps with the
95\% excluded region explored by the $\tau$ channel.

\begin{figure}[t]
\centering
\includegraphics{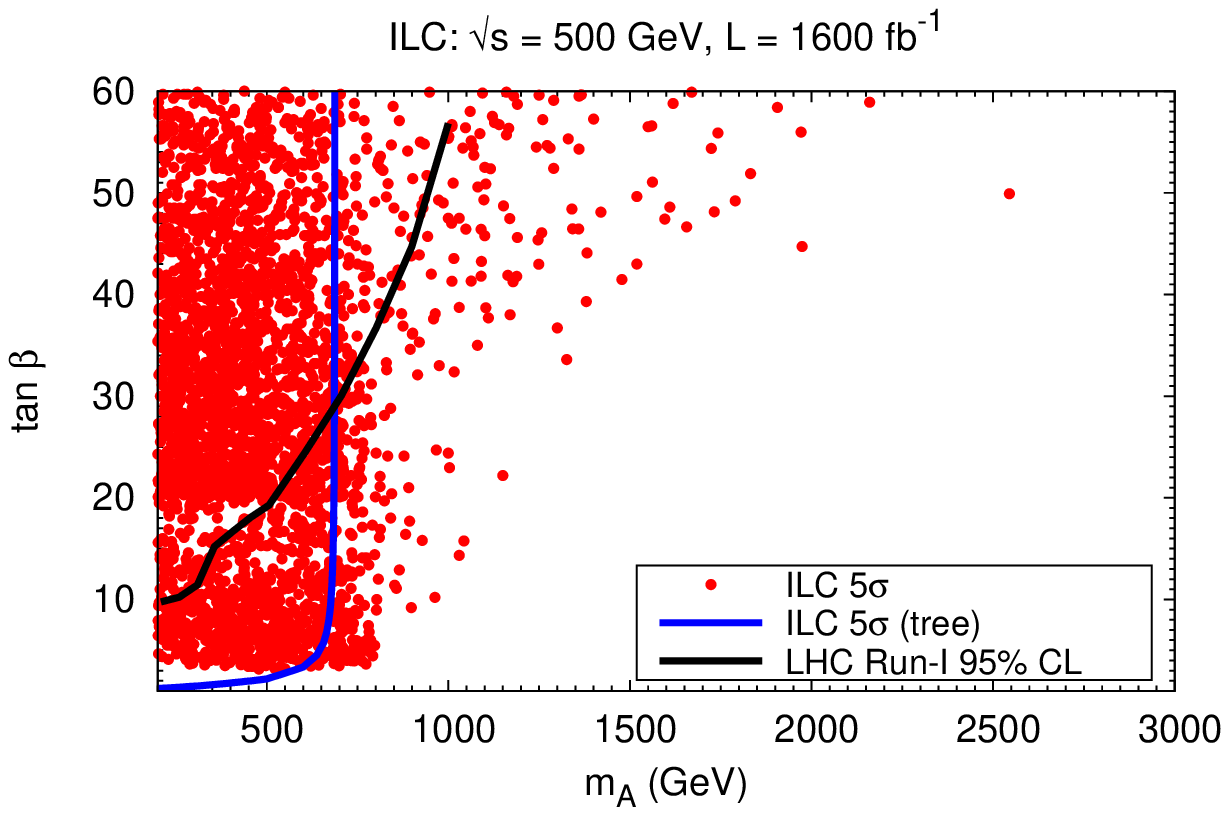}
\caption{\footnotesize Indirect $5 \sigma$ discovery reach of the
  heavy Higgs bosons through the measurement of $R$ at the
  luminosity-upgraded ILC stage with $\sqrt{s}=500~{\rm GeV}$ and
  $L=1600~{\rm fb}^{-1}$ in the ($m_A, \tan \beta$) plane (red
  points).  The others are same as in Fig.1.}
\label{fig:ILC500L}
\end{figure}

\begin{figure}[t]
\centering
\includegraphics{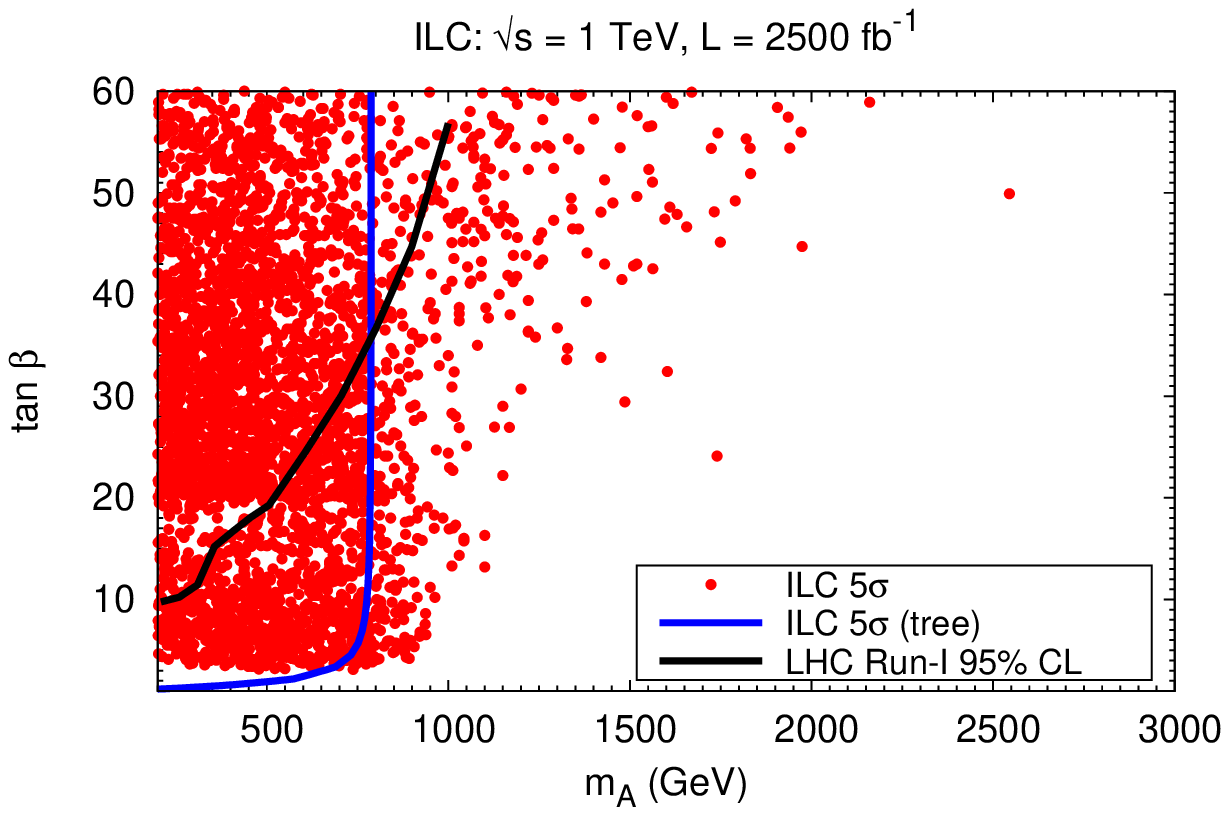}
\caption{\footnotesize Indirect $5 \sigma$ discovery reach of the
  heavy Higgs bosons through the measurement of $R$ at the
  luminosity-upgraded ILC stage with $\sqrt{s}=1~{\rm TeV}$ and
  $L=2500~{\rm fb}^{-1}$ in the ($m_A, \tan \beta$) plane (red
  points).  The others are same as in Fig.1.}
\label{fig:ILC1000L}
\end{figure}

The $5 \sigma$ heavy Higgs mass reach expected at the stage with
$\sqrt{s}=500$ GeV and $L=1600$ fb$^{-1}$ and at the stage with
$\sqrt{s}=1$ TeV and $L=2500$ fb$^{-1}$ are shown in
Fig.~\ref{fig:ILC500L} and Fig.~\ref{fig:ILC1000L}, respectively.
These figures show that luminosity-upgraded ILC programs are very
powerful in measuring the Higgs boson branching fractions and thus
sensitive to higher $m_A$ regions \cite{ILCHiggsWhitePaper}.

Before closing, we comment on the parametric uncertainties from the SM
parameters~\cite{parametric}.
Theoretical calculation on ${\rm Br}(h\to b\bar{b})$
suffers from uncertainties on $m_b$ and
$\alpha_s$, and the uncertainties on the calculation limit the reach of MSSM
parameters by experimental measurements~\cite{Desch:2004cu}.
Although current errors on these parameters~\cite{Eidelman:2004wy}
lead to larger
uncertainties than the expected uncertainties at the future ILC
measurements, it is anticipated that future developments on higher order
calculation and lattice calculation overcome this situation until the
experimental programs take place~\cite{Lepage:2014fla}.

%%%%%%%%%%%%%%%%%%%%%%%%%%%%%%%%%%%%%%%%%
%\section{Discussion and Conclusion}
%%%%%%%%%%%%%%%%%%%%%%%%%%%%%%%%%%%%%%%%%

In this letter, we have presented to what extent the heavy additional
MSSM Higgs bosons can be indirectly accessible by using precision
measurements of the cross section times branching ratios for the
discovered Higgs boson at the future ILC experiments.  We have
employed the ratio of the Higgs boson decay branching fractions ${\rm
  Br}(h\to b \bar{b})/{\rm Br}(h\to WW)$ as a simple but useful
observable for indirectly probing the existence of the heavy Higgs
bosons because this ratio is obtained by using the cross section times
branching fractions, which cancel out the uncertainties of the
production cross section and total decay width of the Higgs boson.  We
have computed the ratio $R$, which can substantially deviate from
unity even for large $m_A$ regions due to large non-decoupling effects
of superparticles, and have exhibited the indirect $5\sigma$ discovery
potential of the additional MSSM Higgs bosons at several planned
stages of the future ILC experiments.  Our analysis shows that the
ILC, in particular with high luminosity, cover many large $m_A$
scenarios in the MSSM.

Last but not least, fingerprinting of Higgs boson couplings at the
loop level including our analysis improves capability of exploring and
distinguishing the MSSM from other models of the electroweak symmetry
breaking, and should be encouraged.

\begin{acknowledgments}
  The authors would like to thank Hiroyuki Taniguchi for early stage
  collaboration.  The work of M. Kakizaki was supported in part by
  Grant-in-Aid for Scientific Research, No. 26104702.  The work of
  S.K. was supported in part by Grant-in-Aid for Scientific Research,
  Nos. 24340046, 23104006 and 30181308.  The work of M. Kikuchi was
  supported in part by Japan Society for the Promotion of Science,
  No. $25\cdot 10031$.  The work of T.M. was supported in part by the
  Sasakawa Scientific Research Grant from The Japan Science Society.
\end{acknowledgments}

\end{document}